\documentclass[12pt]{article}
\input xy
\xyoption{all}

\usepackage{amsmath}
\usepackage{amssymb}

\begin{document}

\vspace*{0.5in}

\begin{center}

{\large\bf A few Ricci-flat stacks as phases of exotic GLSM's}

\vspace*{0.2in}

Eric Sharpe \\
Department of Physics \\
Robeson Hall, 0435 \\
850 Drillfield Drive \\
Blacksburg, VA  24061
\\
{\tt ersharpe@vt.edu}

\end{center}

In this letter we follow up recent work of Halverson-Kumar-Morrison on
some exotic examples of gauged linear sigma models (GLSM's).  
Specifically, they describe a set of
$U(1) \times {\mathbb Z}_2$ GLSM's
with superpotentials that are
quadratic in $p$ fields rather than linear as is typically the
case.  These theories
RG flow to sigma models on branched double covers, where the
double cover is realized via a ${\mathbb Z}_2$ gerbe.  For that gerbe
structure, and hence the double cover, the ${\mathbb Z}_2$ factor in the
gauge group is essential.  In this letter we propose an analogous geometric
understanding of phases without that ${\mathbb Z}_2$, in terms of  
Ricci-flat (but not Calabi-Yau) stacks which look like Fano manifolds with
hypersurfaces of ${\mathbb Z}_2$ orbifolds.

\begin{flushleft}
June 2013
\end{flushleft}

\newpage

\tableofcontents

\section{Introduction}

Over the last few years we have seen a number of advances in our
understanding of gauged linear sigma models (GLSM's),
ranging from non-Kahler compactifications (see {\it e.g.}
\cite{ael,adl,qs1,mqs} and references therein) 
to new developments in Calabi-Yau compactifications,
including realizations of non-complete-intersection target spaces
\cite{ht,ds,hori1,jklmr1},
non-birational
phases \cite{ht,ds,cdhps,bdfik},
nonperturbative realizations of geometry \cite{ht,ds,cdhps,hkm}, 
realizations of noncommutative resolutions \cite{ds,cdhps},
and
localization techniques \cite{benini1,doroud1}.
Those localization techniques have been applied to deduce
new methods for computing Gromov-Witten invariants
\cite{hkm,jklmr2}, as well as many other results, see
{\it e.g.} \cite{gl1,ps,ncgw,hm,gg,beht,bstv}.

This paper will focus on a class of GLSM's described in \cite{hkm}.
These are abelian GLSM's, with a superpotential that is quadratic
in ``$p$'' fields.  Ordinarily, $p$'s appear linearly, acting ultimately
as a form of Lagrange multiplier that forces the semiclassical space
of vacua to be a complete intersection in some space.  Here, one has
a $U(1) \times {\mathbb Z}_2$ gauge theory with a superpotential that
is quadratic in $p$'s, whose large-radius phases are
interpreted by  \cite{hkm}, following \cite{cdhps}, 
to be Calabi-Yau
branched double covers.  The ${\mathbb Z}_2$ gauge factor plays
an essential role, and is ultimately responsible for that branched double
cover structure.

The purpose of this note is to propose an understanding of corresponding
GLSM's without that crucial ${\mathbb Z}_2$.
Omitting it leaves one with a puzzle:
one gets what looks locally
like a single copy of a Fano manifold, which is not Ricci-flat, away
from the former branch locus.  
However, 
there is a unique stack structure one
can impose on that Fano manifold along what would have been the branch locus,
such that the resultant stack is Ricci-flat (but not Calabi-Yau).
We propose that that Ricci-flat stack is the correct 
understanding of the GLSM phase.   

In section~\ref{sect:review} we briefly review the pertinent $U(1) \times
{\mathbb Z}_2$ GLSM's of Halverson-Kumar-Morrison \cite{hkm},
and in section~\ref{sect:proposal} we describe our proposal for the
interpretation of the GLSM's when the ${\mathbb Z}_2$ is omitted.
We include a discussion of general properties of such examples.
These stacks are closely parallel to Enriques surfaces,
and we discuss general aspects of associated superconformal
field theories.  After untangling a potential confusion regarding
covers and quotients in this example, we conclude with a discussion of
the application of the geometric criteria of \cite{hkm} to this case.

\section{Review of GLSM's of Halverson-Kumar-Morrison}
\label{sect:review}

One of the insights of \cite{hkm,joshpriv} was to consider GLSM's in which
the superpotential has terms involving multiple $p$ fields.  Reference
\cite{hkm} tabulates
data for a number of GLSM's of this form
describing Calabi-Yau threefolds at
large-radius, which we have reproduced and extended in 
table~\ref{table:update4}.

Let us briefly outline the analysis of two pertinent examples in that
table, beginning with the first entry.
This describes a GLSM with
4 chiral superfields $x_i$ of charge +1 and 4 chiral superfields $p_a$
of charge $-1$, with a superpotential of the form
\begin{displaymath}
W \: = \:  \sum_{a,b = 1, \cdots, 4}
p_a p_b A^{ab}(x),
\end{displaymath}
where each $A^{ab}(x)$ is of degree two, and gauge group
$U(1) \times {\mathbb Z}_2$, where the ${\mathbb Z_2}$ acts only
on the $p$ fields.  For $r \gg 0$, the $x_i$ are
not all zero, and map out a ${\mathbb P}^3$.
The superpotential defines a mass matrix for the $p_a$'s,
with nonzero eigenvalues away from the locus $\{ \det A^{ab} = 0 \}$,
of degree $(4)(2) = 8$.

On the face of it, this result is confusing -- although the charges
sum to zero, indicative of a Ricci-flat space,
it looks as if theory will
RG flow to a sigma model on ${\mathbb P}^3$.  Closely related
GLSM's with behavior of this form have been discussed in
\cite{cdhps}.  Briefly, because of the ${\mathbb Z}_2$ factor
in the gauge group,
which acts trivially on the $x$'s (and nontrivially on the $p$'s), 
the semiclassical vacua `double' away from the locus where $p$'s become
massless, and a Berry phase around that locus interleaves the
copies appropriately.  

This is a special case of a much more general story about how
two-dimensional theories see trivially-acting finite gauge groups.
Briefly, although the action on geometry is trivial, physics is
nevertheless able to sense their presence via nonperturbative effects,
and the resulting CFT is equivalent to a sigma model on a disjoint
union of multiple spaces. 
See for example \cite{nr,msx,glsm,hhpsa,hs} for a more detailed discussion,
\cite{ajt3,t1,gt1} for mathematical checks of predictions this makes for 
Gromov-Witten theory,
\cite{ncgw,nick-ed-me} for consistency tests and extensions 
of the results of
\cite{cdhps}, and 
\cite{hori1} (as well, of course, as \cite{hkm})
for further examples of this phenomenon in GLSMs.

In any event, as a result of that trivially-acting ${\mathbb Z}_2$,
we can interpret this GLSM phase 
as a branched double cover of the space of $x$'s,
branched over the locus where some of the $p$'s become massless
(modulo behavior at singular loci we shall discuss momentarily).
Specifically, that means this is a branched double cover of
${\mathbb P}^3$, branched over a degree 8 locus.  

As reviewed in {\it e.g.} \cite{cdhps}, if the degree of the branch locus
is $-2$ times the degree of the canonical bundle, the branched double cover
is Calabi-Yau, which implies that this as well as
all of the other examples in 
table~\ref{table:update4} are Calabi-Yau, resolving our puzzle.

Strictly speaking, we should be slightly careful about the description
as a branched double cover.
There is a subtlety that sometimes the branched double cover
is mathematically singular, 
but the GLSM behaves as if it were smooth.  The condition for the
branched double
cover to be (mathematically) singular is that
the branch locus is singular, {\it i.e.}
there exist places where both the determinant and its derivatives vanish::
\begin{displaymath}
\det A^{ab} \: = \: 0, \: \: \:
\frac{\partial}{\partial x_i} \det A^{ab}(x) \: = \: 0.
\end{displaymath}
The condition for the GLSM to be singular are that both $A^{ab}$ and its
derivatives admit the same zero eigenvector:
\begin{displaymath}
A^{ab} p_b \: = \: 0, \: \: \:
p_a \frac{ \partial A^{ab}(x) }{\partial x_i} p_b \: = \: 0,
\end{displaymath}
so that the theory develops a new noncompact branch.
Certainly, whenever the
GLSM is singular, it is trivial to see that the branched double cover
will also be singular.  However, the converse is not true in general --
the branched double cover will often be mathematically singular
even when the GLSM does not develop a new noncompact branch.
This is a generic feature of GLSM's of this form.

This phenomenon was discussed in \cite{cdhps}.  As discussed there, 
in such cases, we believe that physics sees a `noncommutative resolution'
of the branched double cover, as discussed in 
\cite{cdhps,nick-ed-me,ncgw,kuz2}.  For example, this happens in the
present case:  typically there are 80 points on ${\mathbb P}^3$ where
the branched double cover is mathematically singular, and
(based on previous experience in closely related cases) 
we believe the GLSM describes
a noncommutative resolution at those points. 
Given that the phenomenon is generic, we suspect that similar statements
are true of the other entries in table~\ref{table:update4}.  As this
matter is not central to our purpose in this letter, we shall move on. 

In passing, the geometry in the $r \ll 0$ limit of the GLSM is of the same
form.  Here, the $p$'s are not all zero, and define a ${\mathbb P}^3$;
the superpotential defines a mass matrix for $x$'s; the $U(1) \times
{\mathbb Z}_2$ can be reorganized so that the ${\mathbb Z}_2$ acts only
on $x$'s, not on $p$'s; the resulting geometry is again a branched
double cover of ${\mathbb P}^3$, branched along a degree 8 locus
(or a noncommutative resolution thereof).

\begin{table}
\centering
\begin{tabular}{c|cc|c|cc}
$n$ & $n_P$ & $P$-field charges & Exponents in W & 
{\begin{tabular}{c}
(nc res'n of) \\
branched 2-cover of: \end{tabular}}
& {\begin{tabular}{c}
Branch locus \\ degree \end{tabular}} \\ \hline
4 & 4 & (-1,-1,-1,-1) & (2,2,2,2) & 
${\mathbb P}^3$ & 8\\ \hline
4 & 3 & (-2,-1,-1) & (2,2,2) & ${\mathbb P}^3$ & 8\\
5 & 4 & (-2,-1,-1,-1) & (1,2,2,2) & ${\mathbb P}^4[2]$ & 6 \\ \hline
4 & 2 & (-3,-1) & (2,2) & ${\mathbb P}^3$ & 8 \\
5 & 3 & (-3,-1,-1) & (1,2,2) & ${\mathbb P}^4[3]$ & 4 \\
4 & 2 & (-2,-2) & (2,2) & ${\mathbb P}^3$ & 8 \\
5 & 3 & (-2,-2,-1) & (1,2,2) & ${\mathbb P}^4[2]$ & 6 \\
6 & 4 & (-2,-2,-1,-1) & (1,1,2,2) & ${\mathbb P}^5[2,2]$ & 4 \\ \hline
4* & 1 & (-4) & (2) & ${\mathbb P}^3$ & 8\\
5 & 2 & (-1,-4) & (2,1) & ${\mathbb P}^4[4]$ & 2 \\
5 & 2 & (-2,-3) & (2,1) & ${\mathbb P}^4[3]$ & 4 \\
5 & 2 &  (-2,-3) & (1,2) & ${\mathbb P}^4[2]$ & 6 \\
6 & 3 & (-2,-2,-2) & (1,1,2) & ${\mathbb P}^5[2,2]$ & 4 \\
6 & 3 & (-3,-2,-1) & (1,1,2) & ${\mathbb P}^5[2,3]$ & 2 \\
7 & 4 & (-2,-2,-2,-1) & (1,1,1,2) & ${\mathbb P}^6[2,2,2]$ & 2 \\
\end{tabular}
\caption{Examples from \cite{hkm}[table 4].  
Listed are data for GLSM's with gauge
group $U(1) \times {\mathbb Z}_2$, and corresponding interpretations of
$r \gg 0$ phases. The first column counts $x$'s, of charge +1; the next two
columns count $p$'s and give their charges; the fourth column 
gives the exponents
of those $p$'s in the superpotential.  The entry marked with a $*$ is 
described in detail in \cite{hkm}.}
\label{table:update4}
\end{table}

Let us consider one more example, the third entry in the table.
This is a GLSM with gauge group $U(1) \times {\mathbb Z}_2$
as before, 5 chiral superfields $x_i$ of charge +1,
4 chiral superfields $p_a$ of charges -2, -1, -1, -1,
superpotential
\begin{displaymath}
W \: = \: p_1 f_2(x) \: + \: \sum_{a,b=2,3,4} p_a p_b A^{ab}(x),
\end{displaymath}
where $p_1$ has charge $-2$, $p_{2,3,4}$ have charge -1,
and $f_2$, $A^{ab}$ are homogeneous of degree 2.
Again let us study the $r \gg 0$ phase.  Here, the $x_i$ are not all zero,
hence form a ${\mathbb P}^4$, and
the $p_1$ term in the superpotential restricts to the hypersurface
$\{ f_2 = 0 \} \subset {\mathbb P}^4$.  The remaining superpotential
terms act as a mass matrix for $p_{2,3,4}$, and their analysis proceeds
as in the last example.  Briefly, one gets a branched double cover of
$\{ f_2 = 0 \}$, branched over the locus 
\begin{displaymath}
\{ \det A = 0 \} \cap \{ f_2 = 0 \}.
\end{displaymath}
It is straightforward to check that this is another Calabi-Yau.

More generally, for a branched double cover of 
${\mathbb P}^n[d_1, d_2, \cdots]$
to be Calabi-Yau, the branch locus must have degree
\begin{displaymath}
2n \: + \: 2 \: - \: 2 \left( \sum d_i \right),
\end{displaymath}
and it is straightforward to check that the remaining examples in 
table~\ref{table:update4} all define Calabi-Yau branched double covers.

\section{Analysis of the $U(1)$ gauge theories, without ${\mathbb Z}_2$'s}
\label{sect:proposal}

\subsection{Basic proposal}

Now, let us modify the examples above.  Instead of taking the gauge group
to be $U(1)\times {\mathbb Z}_2$, let us take the gauge group  
to be just $U(1)$, omitting the ${\mathbb Z}_2$ that played a crucial
role in the analyses above.  We will make a proposal for the IR geometry
in these cases.

Let us return to the first example in table~\ref{table:update4}.
Recall this theory has 4 chiral superfields $x_i$ of charge $+1$ and
4 chiral superfields $p_a$ of charge -1, together with a superpotential
\begin{displaymath}
W \: = \: \sum_{a,b = 1, \cdots, 4}
p_a p_b A^{ab}(x),
\end{displaymath}
where each $A^{ab}(x)$ is of degree two.  For $r \gg 0$, the $x_i$ are not
all zero, hence the superpotential defines a mass matrix for the $p_a$'s,
with nonzero eigenvalues away from the locus $\{ \det A^{ab} = 0\}$,
of degree eight.

Previously, the extra ${\mathbb Z}_2$ acting only on the $p$'s gave rise
to a ${\mathbb Z}_2$ gerbe structure over 
\begin{displaymath}
{\mathbb P}^3 \: - \: \{ \det A^{ab} = 0 \}, 
\end{displaymath}
which physics
sees as a branched double cover, which we denote $X$.  
Without that ${\mathbb Z}_2$ action,
however, there is no gerbe structure, and hence no double cover.
Thus, we know that this theory RG flows to something that is
one copy of ${\mathbb P}^3$ away from the degree 8 locus
$\{ \det A^{ab} = 0 \}$.

Furthermore, we also know that, because the $U(1)$ charges balance, if this
RG flows to a nonlinear sigma model on some sort of geometry, that
geometry must be Ricci-flat.  (As Enriques surfaces can be built from
K3's realized as projective hypersurfaces, by taking suitable finite
quotients, we hesitate to claim that just because $U(1)$ charges balance
the geometry must be Calabi-Yau.)

In the present case, although we cannot directly `see' the structure
along the degree 8 hypersurface, we can uniquely determine it
by demanding that the result admit a Ricci-flat metric.
The structure so determined
is a hypersurface of ${\mathbb Z}_2$ orbifolds \cite{tonypriv}.
Therefore, we propose that this theory RG flows to\footnote{
Or a noncommutative resolution of any singularities in this stack.
As noncommutative
resolutions will play no significant role in what follows, and we
are largely suppressing them in this paper, we will not discuss
them further.
} ${\mathbb P}^3$ with a hypersurface of ${\mathbb Z}_2$ orbifolds
along
the degree 8 locus $\{ \det A^{ab} = 0 \}$.

This proposal extends in the obvious way to the other examples in
table~\ref{table:update4}.  When the ${\mathbb Z}_2$ gauge symmetry
is omitted, we propose that each RG flows to a sigma model on
a Fano space (as listed in table~\ref{table:update4}) with a hypersurface
of ${\mathbb Z}_2$ orbifolds, (or a noncommutative resolution thereof),
along what was formerly the branch locus.
The resulting stacks are Ricci-flat, though not Calabi-Yau, as we
discuss in the next section.

\subsection{General properties of such examples}

More generally,
Fano spaces $B$ with ${\mathbb Z}_2$ orbifolds along a hypersurface
of degree $-2$ times the canonical class of $B$ admit Ricci-flat metrics.
For example, ${\mathbb P}^n[d_1, d_2, \cdots]$ with
${\mathbb Z}_2$ orbifolds along a hypersurface of degree
\begin{displaymath}
2n \: + \: 2 \: - \: 2 \left( \sum d_i \right)
\end{displaymath}
admit Ricci flat metrics.  These stacks have a simple relationship
to the Calabi-Yau branched double covers branched over the same locus that
were
described earlier in this paper:  they can be
obtained by a global ${\mathbb Z}_2$ orbifold that exchanges the
two sheets of the cover.  The Ricci-flat metric on the Calabi-Yau
branched double cover descends to define a Ricci-flat metric on the stack.
Away from the branch locus, this orbifold
acts effectively and reduces the double cover
to just a single copy of the original space.  At the branch locus, this results
in a ${\mathbb Z}_2$ orbifold structure.

On a variety, codimension one orbifolds are invisible:
${\mathbb C}/{\mathbb Z}_2 = {\mathbb C}$, essentially because if
${\mathbb Z}_2$ acts on $x$ by sign flips, then
\begin{displaymath} 
{\mathbb C}[x]^{{\mathbb Z}_2} \: = \: {\mathbb C}[x].
\end{displaymath}
Here, however, we do not have a space, we have a stack, and stacks can and do
keep track of codimenion one (and even codimension zero) orbifold structures.

Since these stacks have hypersurfaces of ${\mathbb Z}_2$ orbifolds,
locally they look like $[ {\mathbb C} / {\mathbb Z}_2 ] \times
{\mathbb C}^r$ for some $r$.  Such affine stacks have been discussed in the
context of nonsupersymmetric strings (see {\it e.g.} \cite{aps,hkmm}), 
and a number of exotic
behaviors have been described that arise after one deforms by a relevant
operator, by giving a vev to a tachyon.  Here, we are not turning
on a tachyon vev, we are not making a relevant deformation, we are
merely sitting on a Ricci-flat stack.  Moreover, as our stacks are obtained
by a global ${\mathbb Z}_2$ orbifold, there exists a quantum symmetry 
which prevents
twisted sector relevant operators from acquiring a vev and driving
RG flow to a trivial endpoint \cite{allanpriv}.

Although these stacks are Ricci-flat, they are not Calabi-Yau, their
canonical classes are 2-torsion \cite{tonypriv}, just like an Enriques
surface.  In fact, SCFT's for Enriques surfaces and for manifolds
with codimension one ${\mathbb Z}_2$ orbifolds are closely parallel.
For example, we can understand the presence of a Ricci-flat metric on
both as arising because they are ${\mathbb Z}_2$ orbifolds of
Ricci-flat spaces.  Specifically, the codimension one ${\mathbb Z}_2$
orbifolds are ${\mathbb Z}_2$ orbifolds of (Ricci-flat) branched
double covers, and Enriques surfaces are ${\mathbb Z}_2$ orbifolds of
(Ricci-flat) K3 surfaces.  In both cases, the metric on the cover descends.

We can see more or less explicitly that the ${\mathbb Z}_2$ orbifold
of a branched double cover will yield a space with 2-torsion canonical
divisor, by considering a simple example.
Start with an elliptic curve
described as a branched double cover of ${\mathbb P}^1$, branched over
a degree 4 locus (4 points).  If we describe this branched double cover
as solutions to
\begin{displaymath}
y^2 \: = \: x (x-1) (x-\lambda),
\end{displaymath}
(where the two possible square roots for $y$ are the two sheets of the
cover, branched over $x=0, 1, \lambda, \infty$), 
then it is a standard result that
the holomorphic top-form on this elliptic curve can be
written in local coordinates as
\begin{displaymath}
\omega \: = \: \frac{dx}{y}.
\end{displaymath}
The global ${\mathbb Z}_2$ orbifold that exchanges the sheets of the
double cover, acts by sending $y \mapsto -y$ and leaving $x$ invariant.
From the expression above, it should be clear that this does not leave
the holomorphic top-form invariant.  Taking the stack quotient results
in a stacky curve that looks like ${\mathbb P}^1$ with 4 ${\mathbb Z}_2$
orbifold points, which admits a Ricci-flat metric but has 2-torsion canonical
class.

More globally, (2,2) SCFT's can be defined at string tree level
for K\"ahler target spaces and stacks whose canonical
divisors are 2-torsion.  Their Ricci-flatness 
implies the existence of a nontrivial
IR fixed point; K\"ahler implies (2,2) worldsheet supersymmetry.
They do not define spacetime supersymmetric theories, however.
For example,
at string one-loop, one runs into subtleties,
due to the fact that such spaces and stacks are not Spin.  
As a result, the (R,NS)
and (NS,R) sector Fock vacua cannot be defined, as they would describe
spinors on the space.  We can see this problem explicitly in examples,
as follows:
\begin{itemize}
\item In the orbifold $[{\mathbb C}/{\mathbb Z}_2]$, the Fock vacuum
is not defined in the (R,NS) and (NS,R) sectors of the orbifold.
In both cases, because there is a periodic complex fermion $\psi$,
there are two Fock vacua, related by the fermi zero mode:
\begin{displaymath}
| 0 \rangle' \: = \: \psi_0 | 0 \rangle.
\end{displaymath}
As a result, the action of the ${\mathbb Z}_2$ orbifold group on the
two Fock vacua differs by a sign, so following the standard procedure,
under the generator of the ${\mathbb Z}_2$, each Fock vacuum is
multiplied by $\pm \sqrt{-1} = \pm i$.  However, that does not square
to $+1$, and so does not faithfully represent the ${\mathbb Z}_2$
orbifold group.  Such states would break the projection operator
implicit in the orbifold one-loop partition function, and so we see
that the orbifold group action is not well-defined on the Fock vacua
or other states\footnote{
As other states are obtained by multiplying the Fock vacua by perturbative
modes, which clearly do have a well-defined action, failure of the Fock
vacua to have a well-defined action implies that no other states
can have well-defined actions either.}
of (R,NS) and (NS,R) sectors.  See \cite{aps}[section 2] for a related
discussion.
\item For a sigma model directly on an Enriques surface, the fact that
Enriques surfaces do not have spinors means that the Fock vacua in the
(R,NS) and (NS,R) sectors cannot be defined, and hence those sectors
do not exist.
\item If we build a sigma model on an Enriques indirectly, by taking
a ${\mathbb Z}_2$ orbifold of a (2,2) SCFT for a K3 surfaces,
we run into a problem making sense of the ${\mathbb Z}_2$ orbifold,
ultimately for the same reason as in the $[{\mathbb C}/{\mathbb Z}_2]$
orbifold.  On a Spin Kahler manifold $M$, spinors are of the form
\cite{lawson-m}
\begin{displaymath}
\wedge^{\bullet} TM \otimes \sqrt{K_M}.
\end{displaymath}
In the RNS worldsheet formulation, perturbative fermi zero modes realize
the $TM$ factors, and the Fock vacuum transforms as an element of
$\sqrt{K_M}$.  On a Calabi-Yau, this means the Fock vacuum is trivial,
though for open strings on wrapped D-branes, the Fock vacuum couples to
a nontrivial bundle.  In any event, in the present case, the Enriques
${\mathbb Z}_2$ orbifold of a K3 flips the sign of the holomorphic top-form,
which means the Fock vacuum would transform under the ${\mathbb Z}_2$
as $\pm i$, hence the states in (R,NS), (NS,R) sectors of the orbifold are
not well-defined.
\end{itemize}
Thus, one cannot define a type II string
compactification on such spaces and stacks.  However, although the
(R,NS) and (NS,R) sectors are not defined, the (R,R) sector is well-defined,
as the square of the canonical bundle is trivial, and so products of
pairs of spinors can be defined, a statement which has simple specializations
to each of the cases above.  As a result, although type II strings are
not well-defined, a type 0 string compactification on spaces or stacks of
the form above should be well-defined.  (See {\it e.g.} \cite{pol2}chapter 10.6]
for an efficient discussion of type 0 strings.)

In passing, the closed string B model on Enriques surfaces was
discussed in \cite{s-b}, which observed that consistency only required
$K_X^{\otimes 2} = {\cal O}_X$, instead of the full Calabi-Yau condition.
The closed string B model is similarly well-defined on the
the codimension-one ${\mathbb Z}_2$ orbifolds
we are discussing here.

\subsection{Covers and quotients}

To review, given a construction of a branched double cover $X$ from a 
GLSM with gauge group $U(1) \times {\mathbb Z}_2$, we propose that
the corresponding GLSM with smaller gauge group $U(1)$ describes
a ${\mathbb Z}_2$ orbifold of $X$, which we denote $Y$.  
Given that the second theory is
built without quotienting a ${\mathbb Z}_2$, one would have expected that the
result would be a cover, not a quotient.  Locally, we believe that this is
an example of an old CFT phenomenon, that says, roughly, that orbifolding
twice returns the original theory.

Specifically, the space $X$ arises from a GLSM that describes a 
${\mathbb Z}_2$ gerbe over most of ${\mathbb P}^3$, {\it i.e.}
locally\footnote{In an analytic sense, not in the Zariski topology.
Thoughout this paper, we use `local' to mean in an analytic sense.
} $[{\mathbb C}^3/{\mathbb Z}_2]$ where the ${\mathbb Z}_2$ acts
trivially.  The resulting local ${\mathbb Z}_2$ orbifold has dimension
zero
operators in each of two twisted sectors, which ultimately is the reason for
the double cover structure.  Now, that local ${\mathbb Z}_2$ orbifold
has a ${\mathbb Z}_2$ quantum symmetry.  If we construct $Y$ as 
$[X/{\mathbb Z}_2]$, then the ${\mathbb Z}_2$ exchanging the sheets of
the cover is acting by the quantum symmetry\footnote{
The two dimension-zero operators are used to form projection operators on
each of two components.  Acting by the quantum symmetry puts phases
on twisted sectors, which has the effect of exchanging the two projection
operators.
}.  It is a standard result \cite{ginsparg}[section 8.5]
that orbifolding by a cyclic group twice, the second time by the
quantum symmetry produced by the first orbifold, reproduces the original theory.
As a result, this is why, locally, $Y$ can be both a ${\mathbb Z}_2$
orbifold and a two-fold cover of $X$.  

Locally, the picture above gives some reasonable intuition, but we should
also note that the global story is more complicated.  The essential point
is that  
for $X$ to be a branched double cover, the associated gerbe on
${\mathbb P}^3$ is a nontrivial gerbe away from the branch locus.  However,
if we were to globally quotient $Y$ itself (and not a cover, as the GLSM
does) by a trivially-acting ${\mathbb Z}_2$,
the result would be a trivial gerbe away from the branch locus, which could
not yield $X$.  Thus, our intuitive picture of successive orbifolds
only can work locally, not globally, on the spaces themselves.  
(In the UV, the ${\mathbb Z}_2$ actions on the covering spaces may
provide a cleaner description in terms of successive orbifolds, modulo the
fact that one would not be working in a CFT; we will not pursue this
here.)
Nevertheless, old results on successive orbifolds do give some
intuitive explanation for the consistency of this proposal.

\subsection{Geometric criteria}

Finally, let us comment on the geometric criteria in \cite{hkm}.
Table 1 of that reference lists a set of criteria for a GLSM phase to have
a geometric interpretation.  The third entry requires that the R-charge
of every gauge-invariant chiral operator is even.  This is satisfied
by the $U(1) \times {\mathbb Z}_2$ gauge theories, as they remark,
but is not satisfied in the present examples, where one only has a
$U(1)$ gauge theory, as both operators of the form $p^2 f(x)$ 
(of even R-charge) and operators of the form $px$ (odd R-charge,
excluded previously by the ${\mathbb Z}_2$) are allowed.

Clearly, one thing this reflects is the fact that the $U(1)$ gauge theories
we have discussed do not define 
Calabi-Yau's, and do not define spacetime supersymmetric
theories, unlike the $U(1) \times {\mathbb Z}_2$ gauge theories.  

However, we can also gain a more subtle understanding of that condition,
by turning to a different description.
The same condition
also appeared in a different guise as part of the definition of
R-symmetry in matrix factorizations in \cite{nick-ed-me}[section 3.1],
\cite{segal1}.  Those references defined a ${\mathbb C}^{\times}_R$ R-symmetry
to be a ${\mathbb C}^{\times}$ symmetry such that the superpotential has
weight 2, and a ${\mathbb Z}_2 = \{ \pm 1 \} \subset {\mathbb C}^{\times}_R$
acts trivially on the underlying space and is the universal
${\mathbb Z}_2^R$, among other things.  The constraint that the
${\mathbb Z}_2^R$ act trivially on the underlying space is the 
sigma model counterpart to the gauge theory statement of \cite{hkm}[table 1]
that gauge-invariant local operators have even R-charge.

The reason for the requirement that the ${\mathbb Z}_2^R$ act
trivially is that in that case,
that R-charge defines gradings on D-branes.
In matrix factorizations, and
open strings more generally, the ${\mathbb Z}_2^R$ distinguishes branes
and antibranes, and when the entire ${\mathbb C}^{\times}$ R-symmetry
acts trivially on the underlying space, the Chan-Paton factors get an
integral grading which partially defines complexes.

Now, let us look at how this R-symmetry appears in some explicit
examples.
In a Landau-Ginzburg
model over the total space $X$ of a vector bundle ${\cal E} \rightarrow B$,
with superpotential of the form $W = p f(x)$, $p$ a fiber coordinate,
$f(x)$ the pullback of a section of ${\cal E}^*$, if we let the 
${\mathbb C}^{\times}_R$ act on $p$ with weight $2$ and
leave $B$ invariant, then the superpotential has weight 2 and the
${\mathbb Z}_2 \subset {\mathbb C}^{\times}$ acts trivially on $X$.

In the examples of the previous section, one has a superpotential of the
form $W = p^2 f(x)$.  If the ${\mathbb C}^{\times}$ R-symmetry acts on
the $p$'s with weight 1, and leaves the $x$'s invariant, then the
superpotential has weight 2.  Because of the ${\mathbb Z}_2$
orbifold $p \mapsto - p$, the universal ${\mathbb Z}_2^R \subset
{\mathbb C}^{\times}_R$ acts trivially on the orbifold (though not
on the covering space of $p$'s).

In the examples in this section, where there is no ${\mathbb Z}_2$
factor in the gauge group, the ${\mathbb Z}_2^R \subset {\mathbb C}^{\times}_R$
does not act trivially on the underlying stack.  Thus, for example, there
would be no way to distinguish branes from antibranes in this theory.

However, because the present theory seems to be describing a space with
2-torsion canonical divisor, there is for example no open string sector
in the B model \cite{s-b}.  We therefore view the fact that the
${\mathbb Z}_2^R$ acts nontrivially on the scalars
(equivalently, that there are local gauge-invariant operators of odd
R-charge), as a reflection of difficulties defining the open string
sector.

\section{Conclusions}

In this paper, after briefly reviewing some exotic GLSM's described
recently in \cite{hkm}, we proposed a description of the IR limits of
some closely related GLSM's, in terms of Ricci-flat (but not Calabi-Yau)
stacks of the form of Fano spaces with ${\mathbb Z}_2$ orbifolds along
a hypersurface of suitable degree.
We studied properties of SCFT's associated to such stacks and to Enriques
surfaces, which are closely related, as well as some subtleties involving
covers versus quotients in repeated orbifoldings, and concluded with
a discussion of the application of the geometric criteria listed in \cite{hkm}
to these Ricci-flat stacks.

\section{Acknowledgements}

We would like to thank A.~Adams, N.~Addington, J.~Lapan, and
especially I.~Melnikov for discussions of Enriques surfaces at string one-loop
and 
T.~Pantev for many discussions of stacks over many years.
This work was partially supported by NSF grant PHY-1068725.

\end{document}